\documentclass[12pt,preprint]{aastex}

\usepackage{epsfig}
\usepackage{color}

\slugcomment{Accepted for publication in ApJ Letters}
\shorttitle{The hybrid CONe WD + He star scenario for the progenitors of SNe Ia}
\shortauthors{WANG et al.}
\begin{document}

\title{The hybrid CONe WD + He star scenario for the progenitors of type Ia supernovae}

\author{  B. Wang\altaffilmark{1,2},
          X. Meng\altaffilmark{1,2},
          D.-D. Liu\altaffilmark{1,2},
          Z.-W. Liu\altaffilmark{3},
          Z. Han\altaffilmark{1,2}
                    }

\altaffiltext{1}{Yunnan Observatories, Chinese Academy of Sciences, Kunming 650011, China; wangbo@ynao.ac.cn}
\altaffiltext{2}{Key Laboratory for the Structure and Evolution of Celestial Objects, Chinese Academy of Sciences, Kunming 650011, China}
\altaffiltext{3}{Argelander-Institut f\"{u}r Astronomie, Auf dem H\"{u}gel 71, D-53121, Bonn, Germany}

\begin{abstract}
The hybrid CONe white dwarfs (WDs) have been suggested to be possible progenitors of
type Ia supernovae (SNe Ia). In this article, we systematically studied the
hybrid CONe WD + He star scenario for the progenitors of SNe Ia, in which
a hybrid CONe WD increases its mass to the Chandrasekhar mass limit by accreting
He-rich material from a non-degenerate He star. According to a series of detailed
binary population synthesis simulations, we obtained the SN Ia birthrates and delay
times for this scenario. The SN Ia birthrates for this scenario are
$\sim$0.033$-$0.539$\times$$10^{-3}\,{\rm yr}^{-1}$, which roughly accounts for
1$-$18\%  of all SNe Ia. The estimated delay times are $\sim$28\,Myr$-$178\,Myr,
which are the youngest SNe Ia predicted by any progenitor model so far. We suggest that
SNe Ia from this scenario may provide an alternative explanation of type Iax SNe.
We also presented some properties of the donors at the point when the WDs reach the Chandrasekhar mass.
These properties may be a good starting point for investigating the surviving companions of
SNe Ia, and for constraining the progenitor scenario studied in this work.

\end{abstract}


\section{Introduction} \label{1. Introduction}

Type Ia supernovae (SNe Ia) have a prominent role in modern
astrophysics and are the best standard candles for probing the
Universe on cosmological scales due to their high luminosities and remarkable
uniformities (e.g., Riess et al. 1998; Perlmutter et al. 1999).
However, the identity of their progenitors and the physics of their explosion mechanisms
are still uncertain (see Hillebrandt \& Niemeyer 2000; Podsiadlowski et al.\ 2008;
Wang \& Han 2012; Maoz et al. 2014).

SNe~Ia are thought to be thermonuclear explosions of carbon--oxygen white
dwarfs (CO WDs) at about the Chandrasekhar mass, though the means by which
they grow to about the Chandrasekhar mass still remain unclear (see Nomoto et al. 1997).
Two kinds of progenitor models have been proposed as possible mechanisms by
which SNe Ia can be produced, which are the single-degenerate and double-degenerate
models. In the single-degenerate model, a CO WD can accrete H- or He-rich matter
from a non-degenerate star to increase its mass to approach the
Chandrasekhar mass, and then generate a thermonuclear explosion to become an SN Ia, in which
the donor star could be a main-sequence star, a subgiant, a red giant,
or a He star (e.g., Hachisu et al.\ 1996; Li \& van den Heuvel 1997; Langer et al.\ 2000;
Han \& Podsiadlowski 2004; Meng et al. 2009; Wang et al. 2009a).
In the double-degenerate model, SNe Ia arise from the merging of two CO WDs in a close binary.
The closeness of two WDs is due to common envelope evolution, which then enables gravitational
wave radiation to drive orbital inspiral to merger (e.g., Webbink 1984; Iben \& Tutukov 1984).
Some variants of these two models have been proposed to explain the observed diversity of SNe Ia
(for recent reviews see Wang \& Han 2012; Maoz et al. 2014).

According to hydrodynamic simulations, Denissenkov et al. (2013) recently
suggested that convective boundary mixing in an super-AGB star
can prevent the carbon burning from reaching the center,\footnote{Convective
boundary mixing (CBM) flattens a carbon abundance profile below the formal boundary
of the carbon convective shell defined by the Schwarzschild criterion because the CBM
transports carbon from there into the convective shell, where it is burnt. For the
carbon shell burning to steadily propagate to the center, the carbon abundance should
be sufficiently high immediately below the Schwarzschild boundary, i.e., it should
steeply increase with a distance from the boundary. This condition is not fulfilled
in the presence of the CBM. In other words, the CBM actually deprives the carbon shell
burning of its fuel on its way to the center (see Denissenkov et al. 2013).}
and will lead to the formation of a hybrid CONe WD after the star has lost its envelope;
such a WD  has an unburnt CO-core surrounded by a thick ONe zone.
Following the work of Denissenkov et al. (2013), Chen et al. (2014) found that, considering the
uncertainty of the carbon burning rate and the treatment of convective boundaries, hybrid WDs may be
produced even by stars with initial mass $>$$7.0\,M_{\odot}$; the mass of these WDs could be close to
1.3\,$M_{\odot}$ in the extreme case of adopting a factor of carbon burning rate of
0.1.\footnote{The fiducial carbon burning rate is a factor of 1.0 (see Caughlan \& Fowler 1988).
For more detailed discussions on the carbon burning rate see Bennett et al. (2012) and Pignatari et al. (2013).}
It is easy for these hybrid WDs to grow to the Chandrasekhar mass limit by accreting matter,
which could increase the birthrates of SNe Ia if CONe WDs can actually produce SNe Ia. Note that
Denissenkov et al. (2014) recently found that hybrid WDs could reach a state of explosive
carbon ignition, though depending on the convective Urca process and some mixing assumptions.

Motivated by the work of Chen et al. (2014), Meng \& Podsiadlowski (2014) recently investigated
the CONe WD + MS scenario of SN Ia progenitors by a detailed binary population synthesis (BPS) method.
However, a CONe WD can also accrete matter from a He star to increase its mass,
and then explode as an SN Ia (this is referred to as the CONe WD + He star scenario in this work).
The purpose of this Letter is to estimate the SN Ia birthrates and delay times in
this scenario. The paper is organized as follows. In Section 2, we describe our
basic assumptions for numerical calculations. We present the results of our
calculations in Section 3. Finally, a discussion and summary are given in Section 4.

\section{Numerical Methods}

In the CONe WD + He star scenario, a CONe WD accretes matter from a He star when it fills its Roche lobe.
The donor star transfers some of its matter to the surface of the WD, which leads to the increase of the WD mass.
If the WD grows up to 1.378$\,M_{\odot}$, we assume that it explodes as an SN Ia.
Based on the optically thick wind model (Hachisu et al. 1996),\footnote{The optically thick wind
model is still controversial (see Langer et al. 2000).} Wang et al. (2009a) have already obtained a dense model
grid leading to SNe Ia with solar metallicity for various initial WD masses except for $1.30\,M_{\odot}$.
Adopting the assumptions of Wang et al.  (2009a), we obtained the initial parameter space leading to SNe Ia for $M_{\rm WD}^{\rm i}$=$1.30\,M_{\odot}$. Figure 1 presents the contours leading to SNe Ia for different initial WD masses.

In order to obtain SN~Ia birthrates and delay times, a series of Monte Carlo simulations
in the BPS approach are performed. For each BPS realization, we used
Hurley's rapid binary evolution code (Hurley et al. 2002) to follow
the evolution of 4$\times$$10^{\rm 7}$ sample binaries.\footnote{We did not
consider super-winds on the evolution of AGB star. If mass loss in the super-wind is
rapid enough, it can drive expansion of the binary orbit (and of the Roche lobe of the WD
progenitor) faster than stellar evolution, preventing Roche-lobe overflow. This could reduce
the SN Ia birthrates studied in this work.} Following the work of Meng \& Podsiadlowski (2014),
we also assumed that, if the mass of a WD is less than the most massive hybrid one shown in
Figure 5 of Chen et al. (2014), and is not a CO WD, then it is a hybrid CONe WD. These binaries
are followed from star formation to the formation of the CONe WD + He star systems based on
three binary evolutionary channels (i.e., \textit{He star, EAGB} and \textit{TPAGB
channels}; see Wang et al. 2009b). If the initial
parameters of a CONe WD + He star system  are
located in the SN Ia production regions in the plane of
initial orbital period and initial companion mass for its specific initial WD mass (Figure 1),
then an SN Ia is assumed to be formed. The factors of the carbon burning rate are set to 0.1, 1 and 10
based on the Figure 5 in Chen et al. (2014).

We conduct eight sets of  Monte Carlo simulations to
examine their influence on the SN Ia birthrates, where we set
the BPS parameters over a reasonable range (see Wang et al. 2009b).
The details of the initial conditions for the BPS simulations are given in Table 1.
A summary of the various given initial conditions is as follows:
(1) Either constant star formation rate (SFR) over the past 14\,Gyrs or,
alternatively, it is modeled as a delta function in the form of a single starburst.
(2) The initial mass function (IMF) is from either Miller \& Scalo (1979, MS79) or Scalo (1986, S86).
(3) A mass-ratio distribution ($n(q)$) that is either constant, rising  or calculated from the case
in which both binary components are chosen randomly and
independently from the IMF (uncorrelated).
(4) All stars are assumed to be members of binaries which have an initially circular orbit.
(5) The distribution of initial orbital separations is assumed to be constant in $\log a$ for wide binary systems, in which $a$ is the orbital separation (e.g., Han et al. 1995).
(6) The standard equations describing energy are used to calculate the output during the common-envelope (CE) phase (e.g., Webbink
1984). Similar to our previous studies (e.g., Wang et al. 2009b),
we use a single free parameter $\alpha_{\rm ce}\lambda$ to describe the CE ejection process, and adopt
three specific values (0.5, 1.0 and 1.5).

\section{Results}

\subsection{Distribution of Initial WD Masses}

Figure 2 shows the distribution of the initial CONe WD masses of
the WD + He star systems that ultimately produce SNe Ia with
different values of $\alpha_{\rm ce}\lambda$. This distribution is given at
the current epoch by assuming an ongoing constant SFR.
From this figure, we can see that a low value of $\alpha_{\rm ce}\lambda$ tends to lead
to higher initial WD masses. This trend can be understood by the \textit{He star
channel} as defined by Wang et al. (2009b), which allows a stable Roche-lobe overflow, leading to form more massive WDs;
a low value of $\alpha_{\rm ce}\lambda$ in our BPS simulations will
increase the fraction of SNe Ia that can be produced by the \textit{He star
channel}, and thus tend to form more massive WDs.
However, we note that WD formation in the \textit{He star
channel} is different from origin
in super-AGB stars as described by Denissenkov et al. (2013); as such, it is unclear whether or not WDs
from the \textit{He star channel} may be hybrid CONe WDs, as we assume.

\subsection{Birthrates and Delay Times of SNe Ia}

According to the eight sets of simulations for the CONe WD + He star scenario,
the estimated SN Ia birthrates are strongly dependent on the choice of the initial conditions, e.g., they
are sensitive to the choice of the CE ejection parameter, carbon burning rate (CBR),
initial mass function and initial mass ratio distribution, etc.
Notably, if we adopt an extreme mass-ratio distribution with uncorrelated component masses (set 8),
the SN Ia birthrate will decrease significantly. This is because most of the donors in this scenario
are not massive, the result of which is  that WDs cannot accrete enough mass to grow up to the Chandrasekhar mass.

In Figure 3, we compare the evolution of SN Ia birthrates for a constant SFR ($3.5\,M_{\rm
\odot}{\rm yr}^{-1}$; left panel) and a single starburst (right panel).
According to our standard model (set 2),  the SN Ia birthrates
are $\sim$0.298$\times$$0^{-3}\,{\rm yr}^{-1}$, which is roughly one tenth of the observed birthrate
($\sim$3$\times$$10^{-3}\ {\rm yr}^{-1}$; Cappellaro \& Turatto 1997).
Even the largest birthrate in our BPS model (set 7) is only a factor of two greater.
This indicates that the CONe WD + He star scenario
can only be responsible for a part of the total SN Ia birthrate (for other SN Ia formation scenarios see Wang \& Han 2012).
We note that SN Ia birthrates will become lower with the decrease of $\alpha_{\rm ce}\lambda$ (see the left panel).
This is because more binaries after the CE ejection may merge with a low $\alpha_{\rm ce}\lambda$.
In addition, the SN Ia birthrates decrease with the CBR factor; a high CBR factor
will result in a small upper mass limit for the CONe WDs, and consequently a low birthrate.

In Figure 3, we also present the delay time distributions of SNe Ia obtained
from a single starburst  (see the right panel).
From this panel, we see that SN Ia explosions occur between $\sim$28\,Myr
and $\sim$178\,Myr after the starburst, which may contribute to
the population of young SNe Ia in late-type galaxies.
Wang et al. (2009b) found that the minimum delay time from the CO WD + He star scenario
is $\sim$45\,Myr, which is longer than the results obtained in this work. It seems
that SNe Ia from the CONe WD + He star scenario are the youngest of all current
progenitor models.

\subsection{Surviving Companions of SNe Ia}
The donor star in the CONe WD + He star scenario would survive and potentially
be identifiable if the WD is completely disrupted at the moment of SN explosion (e.g., Wang \& Han 2009; Pan et al. 2010; Liu et al. 2013).
By interpolating in the three-dimensional grid (initial WD mass,
initial companion mass and initial orbital period) of the WD + He star systems (Figure 1),
we can obtain many properties of companions when the WDs
grow to 1.378$\,M_{\odot}$, e.g., the luminosities, the effective temperatures,  the orbital velocities,  the
surface abundances, etc. These properties may be observed and be used to help identify the companions.
Figure 4 shows an example of the distributions of the properties of companions in
the plane of the effective temperature and luminosity at the point when the WD increases its mass
to 1.378$\,M_{\odot}$, which may be helpful for identifying the surviving companions.
In this figure, we also present the final region that is obtained from the binary
calculations in Figure 1 (see the dashed line). The possible He companion star in
the SN 2012Z progenitor system is located in this region (for a discussion see Section 4).

\section{Discussion and Conclusions}

SNe Ia from the CONe WD + He star scenario may exhibit some special properties.
In this scenario, the WD accretes material from a non-degenerate He star,
which could result in the detection of He lines in the early spectra of
such SNe Ia. In addition, SNe Ia from this scenario are relatively young and have
delay times as short as $\sim$28\,Myr; such SNe Ia may be detected in galaxies
with recent star formation. Some previous works indicate that the SN Ia
luminosities at maximum could be mainly dependent on  the
carbon abundance, i.e., a low carbon abundance leads to a
smaller amount of $^{\rm 56}{\rm Ni}$ synthesized in the thermonuclear
explosion, which results in a lower peak luminosity of SNe Ia (e.g., Umeda et
al. 1999).
Compared with normal CO WDs, hybrid WDs have a relatively low carbon
abundance (e.g., Denissenkov et al. 2014). Therefore, SNe Ia from these hybrid WDs could be expected to have a lower
peak luminosity, and a lower explosion
energy (a relatively low ejecta velocity could thus be expected).

It has recently been proposed that one sub-class of SNe~Ia
is so distinct as to be classified separately from the bulk
of SNe Ia, with a suggested name of type Iax SNe
(SNe~Iax), which contain SNe resembling the prototype event
SN 2002cx (e.g., Foley et al.\ 2013).
This type of SNe may be excellent candidates for observational
counterparts of SNe Ia via the CONe WD + He star scenario.
SNe~Iax have the maximum luminosities as low as that of the faint 1991bg-like events, and have lower
maximum-light velocities compared with normal ones, but they show iron-rich
spectra at maximum light like the bright 1991T-like objects (Foley et al.\ 2013).
So far, about 25 SNe~Iax have been identified, in which
two of them show strong He lines in their spectra,\footnote{Foley et al.\ (2013)
speculated that all SNe Iax may have significant amounts of He in their ejecta.}
and most of them have been discovered in
late-type galaxies (Foley et al.\ 2013).
Lyman et al. (2013) found that the host population of SNe~Iax is very young, which can be comparable with that of type
IIp SNe, and thus suggested that SNe~Iax may have a delay time of 30$-$50\,Myr.
Foley et al. (2014) recently constrained the
progenitor system of SN 2008ha to have an age of $<$80\,Myr.
The estimated birthrates of SNe~Iax may account for
5$-$30\% of the overall SN Ia birthrate (e.g., Li et al. 2011;
Foley et al. 2013; White et al. 2014). The above observed properties of SNe Iax seem comparable
with those from the CONe WD + He star scenario.

McCully et al. (2014) recently found that one SN Iax (i.e., SN 2012Z)
was probably an explosion of a WD accreting matter from a He star. In Figure 4, we can see that
the the possible He companion star is a little cooler than our BPS results, but this
is merely a selection effect due to the initial conditions of the populations we consider in our
BPS studies; it still lies in the region that can potentially be reached by our binary simulations.
Long period systems in Figure 1 should contribute significantly towards the number of systems in
the vicinity of SN 2012Z, even though it is difficult for our current BPS approach to reflect this.
Thus, we cannot exclude the He donor star as a probable companion of SN 2012Z.

However, the CONe WD + He star scenario cannot explain one particular SN Iax, i.e, SN 2008ge,
which was discovered in an old environment, hosted by an S0
galaxy with no massive stars nor any sign of star formation (Foley et al.\ 2010).
This indicates that SNe Iax have a heterogenous class of progenitors.
We note that some other models have already been  proposed to produce SNe Iax, e.g.,
a failed deflagration model of Chandrasekhar mass WD (e.g.,
Jordan et al.\ 2012; Kromer et al.\ 2013; Long et al. 2014), a specific class
of He-ignited WD explosions (Wang et al. 2013),  and the CONe WD + MS scenario (Meng \& Podsiadlowski 2014).

Observationally, some massive WD + He star binaries (e.g.,
HD 49798 with its WD companion and V445 Pup) are candidates of SN Ia progenitors.
(1) HD 49798 is a H depleted subdwarf O6 star that contains
a massive WD companion with an orbital period of
1.548\,d (e.g., Bisscheroux et al. 1997). Mereghetti et al. (2009) obtained the masses of
these two components, in which the  WD mass is
1.28$\pm$0.05$\,M_{\odot}$ and the He star mass is
1.50$\pm$0.05$\,M_{\odot}$.
(2) V445 Pup is a He nova.
The light curve fitting by Kato et al. (2008)
shows that the WD mass is $\gtrsim$$1.35\,M_{\odot}$.
Woudt et al. (2009) deduced that the pre-outburst luminosity of the
system was log$(L/L_{\odot})$=4.34$\pm$0.36, which is compatible
with a 1.2$-$$1.3\,M_{\odot}$ He star that is burning its He shell (see also Piersanti et al. 2014).
Goranskij et al. (2010) recently reported that
the most probable orbital period for this binary is $\sim$0.65\,d.
The parameters of these two binaries are located in the initial-parameter-space contours
for producing SNe Ia (see Figure 1).
Thus, they are possible progenitor candidates of SNe Ia.
However, it is still uncertain which type of WDs in these two binaries are,
e.g., CO WDs, CONe WDs or ONe WDs.
If they are CONe WDs, they could form SNe Ia through
the scenario studied in this work.

By using a detailed  BPS approach and assuming CONe WDs can produce SNe Ia, we systematically investigated the
hybrid CONe WD + He star scenario for the progenitors of SNe Ia. We obtain the
birthrates and delay times for this scenario. The birthrate from this scenario could account for
1$-$18\% of total SNe Ia, the specific proportion of which
is strongly sensitive to uncertainties in some input parameters for the Monte Carlo simulations.
SNe Ia from this scenario could be as young as $\sim$28\,Myr, which are the youngest SNe Ia ever modeled.
We found that SNe Ia from this scenario will exhibit some special properties when compared with normal ones, and may explain some SNe Iax.
We also provided the properties of donors when the WD mass increases to 1.378$\,M_{\odot}$.
These properties are a starting point for investigating the surviving companions of SNe Ia.
In order to set further constraints on the hybrid CONe WD + He star scenario, large samples of massive
WD + He star systems and surviving companions are needed. We hope that our work stimulates numerical
simulations on thermonuclear explosions of hybrid CONe WDs.

\begin{acknowledgements}
We acknowledge the anonymous referee for valuable comments that helped us
to improve the paper. We thank Philipp Podsiadlowski, Pavel A. Denissenkov, Michael C. Chen, Stephen Justham and Jujia Zhang
for their helpful discussions. This work is supported by the 973 Program of
China (No. 2014CB845700), the NSFC (Nos. 11322327, 11103072, 11033008, 11473063 and 11390374),
and the Yunnan Province (Nos. 2013FB083 and 2013HB097).

\end{acknowledgements}

\clearpage

\begin{figure}
\begin{center}
\epsfig{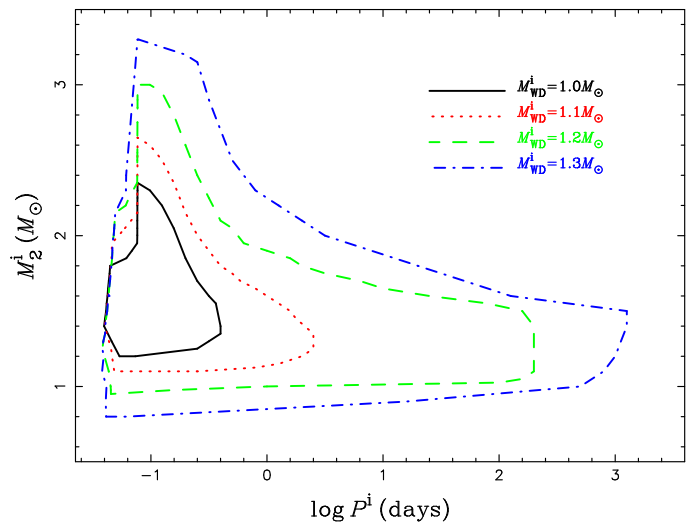}
\caption{Contours in
the initial orbital period and initial companion mass plane for CONe WD binaries that produce
SNe Ia for various initial WD masses.}
\end{center}
\end{figure}

\begin{figure}[]
\begin{center}
\includegraphics[width=10cm,angle=0]{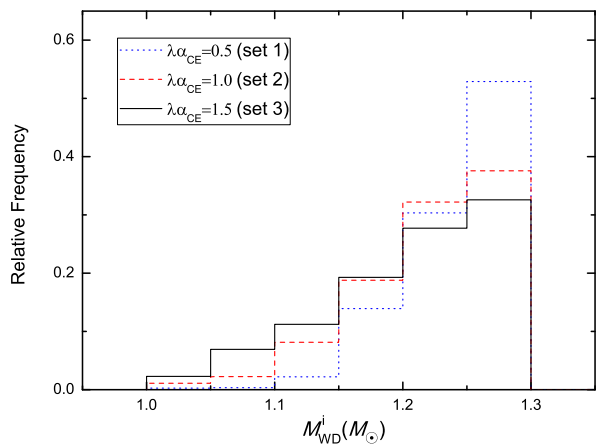}
 \caption{Distribution of the initial CONe WD masses that can ultimately produce SNe Ia with different values of $\alpha_{\rm ce}\lambda$.}
  \end{center}
\end{figure}

\begin{table}
\begin{center}
\caption{SN Ia Birthrates for Different BPS Simulation Sets, in which Set 2 is Our
Standard Model.}
\begin{tabular}{cccccccc}
\hline \hline
Set & $\alpha_{\rm ce}\lambda$ & ${\rm CBR}$ & ${\rm IMF}$ & $n(q)$ & Rate ($10^{-3}$\,yr$^{-1}$)\\
\hline
$1$ & $0.5$ & $0.1$     & ${\rm MS79}$   & ${\rm Constant}$  &  $0.073$\\
$2$ & $1.0$ & $0.1$     & ${\rm MS79}$   & ${\rm Constant}$  &  $0.298$\\
$3$ & $1.5$ & $0.1$     & ${\rm MS79}$   & ${\rm Constant}$  &  $0.473$\\
$4$ & $1.5$ & $1$       & ${\rm MS79}$   & ${\rm Constant}$   & $0.348$\\
$5$ & $1.5$ & $10$      & ${\rm MS79}$   & ${\rm Constant}$   & $0.097$\\
$6$ & $1.5$ & $0.1$     & ${\rm S86}$    & ${\rm Constant}$  &  $0.269$\\
$7$ & $1.5$ & $0.1$     & ${\rm MS79}$   & ${\rm Rising}$    &  $0.539$\\
$8$ & $1.5$ & $0.1$     & ${\rm MS79}$   & ${\rm Uncorrelated}$ & $0.033$\\
\hline
\end{tabular}
\end{center}
\end{table}

\begin{figure*}
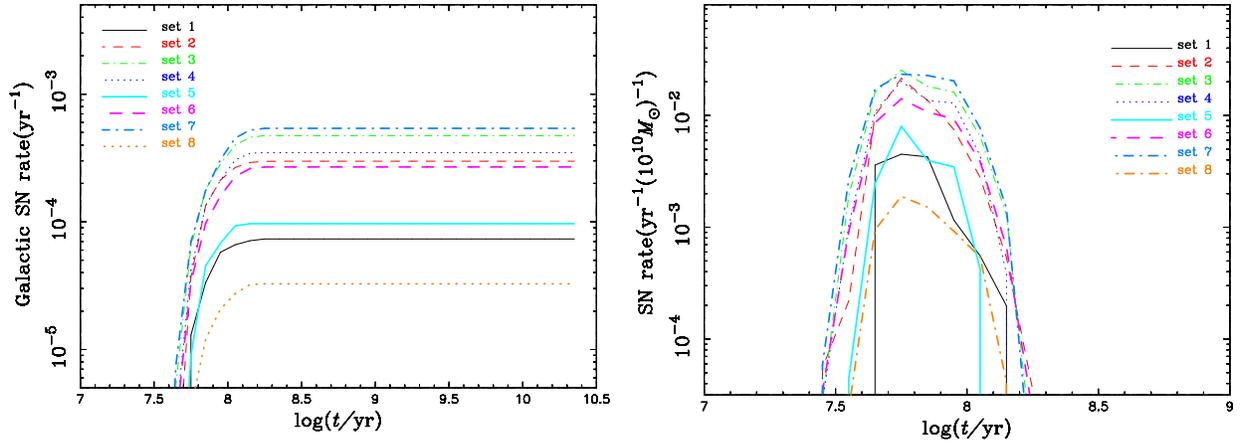

\centerline{\epsfig{file=f3a.ps,angle=270,width=8cm}\ \
\epsfig{file=f3b.ps,angle=270,width=8cm}} \caption{Left panel: the evolution
of SN Ia birthrates for a constant SFR with different
BPS simulation sets. Right panel: similar to the left panel, but for a single starburst.}
\end{figure*}

\begin{figure}
\begin{center}
\includegraphics[width=6.5cm,angle=270]{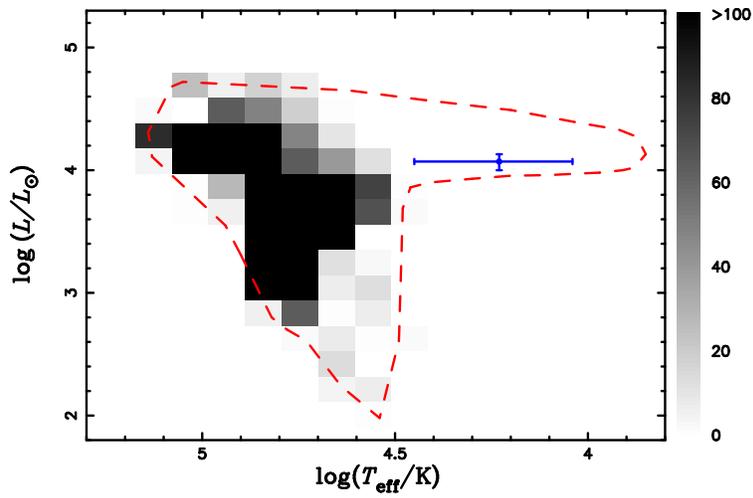}
 \caption{Distribution of properties of the donors in the plane of ($\log T_{\rm eff}$,
$\log L$) when the WDs grow to 1.378$\,M_{\odot}$. Here, we set $\alpha_{\rm ce}\lambda=1.5$ (set 3). The dashed line denotes
the final region obtained from the binary calculations in Fig. 1.
The error bars present the location of the possible companion in the SN 2012Z progenitor system, the luminosity
and temperature of which are based on a black-body approximation of the measurements of McCully et al. (2014).}
\end{center}
\end{figure}

\end{document}